\documentclass{article}

\usepackage{graphicx}
\usepackage{arxiv}
\usepackage[numbers]{natbib}
\usepackage{url}
\usepackage{amsmath, amssymb, amsthm}
\newtheorem{condition}{Condition}[section]

\DeclareMathOperator*{\argmax}{arg\,max}

\DeclareMathOperator{\pr}{P}

\newcommand{\longto}{\longrightarrow}
\DeclareMathOperator{\expit}{expit}

\usepackage{xcolor}

\newtheorem{theorem}{Theorem}

\title{Optimal Dynamic Treatment Regime Estimation in the Presence of Nonadherence}

\author
{Dylan Spicker \\
Department of Mathematics and Statistics, University of New Brunswick, Saint John, New Brunswick, Canada
\And
Michael P. Wallace  \\
Department of Statistics and Actuarial Science, University of Waterloo, Waterloo, Ontario, Canada
\And 
Grace Y. Yi\\
Department of Statistical and Actuarial Sciences, \\
Department of Computer Science, University of Western Ontario, London, Ontario, Canada}

\date{January 12, 2024}

\begin{document}

\maketitle

\begin{abstract}
Dynamic treatment regimes (DTRs) are sequences of functions that formalize the process of precision medicine. DTRs take as input patient information and output treatment recommendations. A major focus of the DTR literature has been on the estimation of optimal DTRs, the sequences of decision rules that result in the best outcome in expectation, across the complete population were they to be applied. While there is a rich literature on optimal DTR estimation, to date there has been minimal consideration of the impacts of nonadherence on these estimation procedures. Nonadherence refers to any process through that an individual's prescribed treatment does not match their true treatment. We explore the impacts of nonadherence and demonstrate that generally, when nonadherence is ignored, suboptimal regimes will be estimated. In light of these findings we propose a method for estimating optimal DTRs in the presence of nonadherence. The resulting estimators are consistent and asymptotically normal, with a double robustness property. Using simulations we demonstrate the reliability of these results, and illustrate comparable performance between the proposed estimation procedure adjusting for the impacts of nonadherence and estimators that are computed on data without nonadherence. 
\end{abstract}

\section{Introduction}
Precision medicine is a field of medical science that aims to tailor treatment recommendations to specific patient characteristics. Precision medicine can be formalized in the longitudinal setting through the use of dynamic treatment regimes (DTRs) \citep{Moodie_DTR_Book,Laber_DTR_Book}. A DTR encodes a sequence of decision rules that take as input patient information, (be that demographic, genetic, health, or other factors), and output a treatment recommendation (among a set of candidate treatment alternatives). Often several treatment decisions need to be made in sequence, where each decision may impact subsequent decisions and where the effects of treatment may be delayed. DTRs capture this sequential decision-making procedure. Many statistical methods for the estimation of \emph{optimal DTRs} have been developed \citep{QLearning,Murphy_2003,AIPW,OWL,dwols,AOL}. We focus on one such technique, \emph{G-estimation} \citep{Robins2004}. G-estimation provides an efficient and robust estimation procedure for optimal DTRs through the use of estimating equation theory.

Methodologies for estimating optimal DTRs have typically required the assumption that all variables are measured without error. Recent work has considered the impact of measurement error in the variables used to inform treatment recommendations, demonstrating that in this setting, traditional methods tend to estimate suboptimal DTRs \citep{Spicker2019_ME}. To the best of our knowledge, no work on optimal DTR estimation has considered the impacts of \emph{nonadherence}. Nonadherence refers to the case where an individual's prescribed treatment recorded in the available data does not correspond to the treatment that they ultimately took. Nonadherence can be viewed as measurement error in the treatment indicators. The lack of methodologies for addressing nonadherence in optimal DTR estimation is of particular concern owing to the prevalence of imperfect adherence in much medical data.

While the impacts of nonadherence have not been addressed for optimal DTR estimation, they have been studied in the broader medical literature. Researchers have studied rates of nonadherence and techniques for improving adherence \citep{Adherence2,Adherence4,Adherence1,Adherence3}. Statistically, any analysis that is conducted using data that are subject to nonadherence, but that ignores the impacts of nonadherence, is referred to as an \emph{intention-to-treat} (ITT) analysis \citep{ITT2,ITT}. In an ITT analysis the causal impact of treatment prescription is estimated, rather than the impact of treatment itself. From a health policy perspective this is a credible rationale for using ITT analyses as clinicians only control the treatments that patients are prescribed, not the ones that they take.

This rationale is particularly well justified when the available alternatives are \emph{as-treated} or \emph{per protocol} analyses \citep{ITT_v_PP}. In an as-treated analysis, the actual treatment is directly used without considering treatment assignment. In a per protocol analysis, only the individuals who adhered to their prescribed treatments are analyzed. Both of these analytical approaches are generally inferior to an ITT analysis as they introduce additional sources of confounding and bias. In contrast, an ITT analysis can often be interpreted as estimating the causal effect of treatment prescription. However, the utility of an ITT analysis depends on assumptions that are often violated in practice \citep{ITTBAD}. The typical justification of an ITT analysis necessitates that the adherence rates in the study population will be equivalent to the adherence rates in the population as a whole. While this assumption may be untenable itself, it also precludes the possibility of influencing adherence rates in the general population. 

To understand whether it is desirable to attempt to improve treatment adherence we must estimate the true treatment efficacy. It is often of direct scientific interest to estimate treatment efficacy. While this cannot be done through an as-treated or per protocol analysis, it is possible to develop treatment efficacy approaches to estimation. We derive a treatment efficacy approach for consistently estimating the optimal DTR in data that are subject to patient nonadherence. Our technique is a modified version of G-estimation. We motivate the work by considering an analysis of data from the Multicenter AIDS Cohort Study (MACS) \citep{MACS}. The study focused on the treatment of HIV/AIDS, and contains biological and behavioral information on over $7000$ men collected from participants every six months. These data have been used in the past to fit optimal DTRs for the timing of treatment interventions \citep{HernanMACS,Wallace_2016}. Starting in 1996, the study began to collect information regarding patient adherence to treatment demonstrating that, for some members of the population, adherence to prescribed therapies is not perfect  \citep{MACS_Adherence}. Our proposed methodologies provide an approach to the analysis of these data that correctly accounts for the observed nonadherence.

\section{Methodology}
\subsection{Optimal Dynamic Treatment Regime Estimation}
We define a $K$-stage DTR as a sequence of functions, denoted $d_j$ for $j=1,\dots,K$. At each stage we take $d_j\colon\mathcal{H}_j\longto\mathcal{A}_j$ to be a function mapping from patient histories, $\mathcal{H}_j$, to possible treatment options, $\mathcal{A}_j$. We start with the assumption that treatments are binary, giving $\mathcal{A}_j = \{0,1\}$ for each $j$. These techniques are extended to arbitrary discrete treatments in Section~\ref{sec::multiple_treatment_alternatives}. The patient history at stage $j$, denoted $H_j \in \mathcal{H}_j$ comprises all the patient information measured at stage $j$, denoted $X_j$, as well as all previous patient information $(X_1,\dots,X_{j-1})$, and all previous treatment assignments, $(A_1,\dots,A_{j-1})$. The history vector, $H_j = (X_1,A_1,X_2,\cdots,A_{j-1},X_j)$, contains all information that is available to inform treatment assignment at stage $j$. Each function, $d_j$, takes in the patient history and outputs a treatment recommendation. Taken together, $d = (d_1,\dots,d_K)$ is called a $K$-stage DTR.

We consider a numeric outcome, denoted $Y$, observed after the final treatment stage, and coded such that larger values are preferable. The goal of optimal DTR estimation is to estimate the DTR that results in the optimal outcome, in expectation, across the population. This can be formalized using the framework of potential outcomes \citep{PotentialOutcomes_3, Robins_1986}. Define the random quantity, $Y^d$, to be the \emph{potential outcome} of $Y$, supposing that $d$ were followed. That is, $Y^d$ is the outcome that would be observed if $A_j = d_j(H_j)$ for all $j=1,\dots,K$. Define $V(d) = E[Y^d]$ to be the \emph{value} of $d$. Then, $d^\text{opt} = \argmax_{d\in\mathcal{D}} V(d)$ is the optimal DTR on $\mathcal{D}$, where $\mathcal{D}$ is the space of all possible $K$-stage DTRs under consideration.

Our proposed methodologies rely on three standard causal identifiability assumptions: \emph{no unmeasured confounding}, also called the \emph{sequential randomization assumption}; the \emph{stable unit treatment value assumption} (SUTVA); and \emph{positivity} \citep{SUTVA,NUC}. Briefly, the no unmeasured confounding assumption requires that any variables that influence both the outcome and treatment assignment are measured in the data. The SUTVA requires that there is only one version of each treatment option and that no individual's treatment assignment impacts any other individual's outcome. Positivity states that, for every individual, $0 < \pr(A_j=1\mid H_j) < 1$. 

Many optimal DTR estimation techniques rely on \emph{backwards induction}. In backwards induction, estimation begins at the final stage of the DTR, and proceeds backwards through the regime. To formalize this procedure, we introduce two quantities. The \emph{$Q$-function} is defined as $Q_{K+1}(h_{K+1},a_{K+1}) = Y$, and, for $j=1,\dots,K$, $Q_j(h_j,a_j) = E[V_{j+1}(h_j, a_j, X_{j+1})\mid H_j=h_j, A_j=a_j]$. Here $V_j$ is the corresponding \emph{$j$th stage value function}, defined as $V_{j}(h_j) = \max\left\{Q_j(h_j, 1), Q_j(h_j, 0)\right\}$, for $j=1,\dots,K+1$. For $j=1,\dots,K$, $Q_j$ represents the expected outcome for an individual with a given observed history, if a particular treatment is assigned. If $Q_j(H_j,1) > Q_j(H_j,0)$, $A_j = 1$ is preferable for an individual with history $H_j$. As a result, $d_j^\text{opt} = I(Q_j(H_j,1) > Q_j(H_j,0))$ is the optimal stage $j$ decision function. The sequential optimization of $Q$-functions defines the optimal DTR \citep{Laber_DTR_Book}.

Taking $A_j\in\{0,1\}$, $Q_j(H_j, A_j) = \nu_j(H_j) + A_jC_j(H_j)$, for some arbitrary functions $\nu_j$ and $C_j$. The $j$th stage value function, $V_j$, will take on the value of $\nu_j(H_j)$, if $C_j(H_j) \leq 0$ and $\nu_j(H_j) + C_j(H_j)$ otherwise. The optimal treatment at each stage is determined solely by $C_j(H_j)$, which we call the \emph{contrast function}. If $C_j(H_j) > 0$ then $A_j^\text{opt} = 1$. If $C_j(h_j)$ is estimated from available data, then $d_j(H_j) = I(C_j(H_j) > 0)$ optimally assigns treatment. The parameter $\psi_j$ indexes the contrast functions, so $Q_j(H_j, A_j) = \nu_j(H_j) + A_jC_j(H_j; \psi_j)$. We assume that the form of $C_j$ is correctly known. We refer to $\nu_j(H_j)$ as the \emph{treatment-free model}, as it captures the impact of the patient history on the outcome not mediated through treatment. The treatment-free model is a nuisance model for optimal DTR estimation.

\subsection{G-Estimation}
G-estimation is a procedure for estimating an optimal DTR using the contrast functions. Suppose that we observe complete individual-level information for $n$ individuals, giving $(X_{i,1}, A_{i,1}, \cdots, X_{i,K}, A_{i,K}, Y_i)$ for each $i=1,\dots,n$. When the individual index $i$ is arbitrary, we omit it. Define $\widetilde{V}_{K+1} = Y$, and, for $j=1,\dots,K$, $\widetilde{V}_{j} = \widetilde{V}_{j+1} + (A_{j}^\text{opt} - A_{j})C_j(H_{j}; \widehat{\psi}_j)$, where $\widehat{\psi}_j$ is an estimator for $\psi_j$. We refer to these as \emph{pseudo outcomes}. If $\widehat{\psi}_j$ is almost surely consistent for $\psi_j$, then $E[\widetilde{V}_{j+1}\mid H_{j},A_{j}] = Q_{j}(H_{j},A_{j}) = \nu_j(H_{j}) + A_{j}C_{j}(H_{j})$ almost surely. Suppose the treatment assignment probabilities, $\pi_j(H_j)\triangleq\pr(A_j = 1\mid H_j)$ are known. \citet{Robins2004} introduces G-estimation for optimal DTR estimation by defining \begin{equation}
    U_j(\psi_j) =  \sum_{i=1}^n \lambda_j(H_{i,j})\left\{A_{i,j} - \pi_j(H_{i,j})\right\}\left\{\widetilde{V}_{i,j+1} - A_{i,j}C_j(H_{i,j};\psi_j) + \theta_j(H_{i,j})\right\}, \label{eq::g_estimation_j}
\end{equation} where $\lambda_j(H_j)$ and $\theta_j(H_j)$ are arbitrary functions of the data, with the dimension of $\lambda_j(H_j)$ matching that of $\psi_j$. Solving $U_j(\psi_j) = 0$ produces an estimator $\widehat{\psi}_j$ of $\psi_j$, that is consistent provided certain regularity conditions hold. Using these estimators to compute the pseudo outcome for stage $j-1$, the process continues to estimate all contrast functions.

When the treatment probabilities are not known, they may be estimated via a parametric model, say $\pi_j(H_j;\alpha_j)$ with a parameter vector $\alpha_j$. Often estimators of $\alpha_j$ are expressible as the solution to a set of unbiased estimating equations, $U_\text{trt, j}(\alpha_j)$. The estimators resulting from stacking $U_\text{trt,j}$ with $U_j$ and jointly solving are consistent if the treatment probability model is correct. Commonly, $\theta_j(H_j)$ will be a specified model for the treatment-free component, indexed by $\beta_j$, $\theta_j(H_j; \beta_j) = -\nu_j(H_j)$. \citet{Robins2004} derives a form for $\lambda_j(H_j)$ for locally efficient estimators. The form for local efficiency is often complex, so it is common to take $\lambda_j(H_j)$ as $(\partial/\partial\psi_j) C_j(H_j;\psi_j)$ \citep{Moodie_DTR_Book,Laber_DTR_Book}. 

When $\theta_j(H_j;\beta_j)$ models $-\nu_j(H_j)$, we will commonly estimate $\beta_j$ using a set of unbiased estimating equations, $U_\text{tf,j}(\beta_j)$. We jointly estimate $\alpha_j$, $\beta_j$, and $\psi_j$ by stacking $U_{\text{trt},j}$, $U_{\text{tf},j}$, and $U_j$. The resulting $\widehat{\psi}_j$ is \emph{doubly robust} in that, if either of the models $\theta_j(H_j;\beta_j)$ or $\pi_j(H_j;\alpha_j)$ are correctly specified, $\widehat{\psi}_j$ is consistent for $\psi_j$, provided certain regularity conditions. This double robustness is a highly desirable property that we preserve in our procedure.

\subsection{Nonadherence in Dynamic Treatment Regimes}
Nonadherence refers to any scenario where the true treatment indicator, $A_j$, is not universally observable. Instead, we may observe $A_j^*$, which we refer to as a \emph{prescribed treatment}, or $A_j^{**}$, which we refer to as a \emph{reported treatment}. The distinction is that $A_j^*$ is an antecedent of $A_j$, while $A_j$ is an antecedent of $A_j^{**}$. A patient is prescribed $A_j^*$, they take $A_j$, and they report that they took $A_j^{**}$. We may observe any subset of $\{A_j^*, A_j, A_j^{**}\}$ for each individual. 

In the context of DTRs, nonadherence has been studied as it relates to the estimation of the value of a regime \citep{Hernan_2006,Cotton_2011,Han_2021}. These techniques characterize $V(d)$ when data are subject to nonadherence, rather than estimating the optimal DTR. For optimal DTR estimation, the causal structure of the problem needs to be considered. When patient histories are measured with error, an analysis ignoring the impacts of error may be justified by viewing the mismeasured variates as predictors themselves \citep{Spicker2019_ME}. This does not materially change the causal question of interest. In the event of nonadherence, however, the causal estimand necessarily changes. When $A_j^*\neq A_j$, then, $A_j^*$ may influence $Y$ both directly and indirectly, through its impact on $A_j$. Taken together, these two effects constitute the ITT effect. If, in place of $A_j^*$, we measure $A_j^{**}$, an analysis ignoring nonadherence and using the reported proxy in place of the true treatment cannot be interpreted causally. This will not give an estimate for the ``causal effect of reported treatment'' owing to unmeasured confounding. In this analysis the true treatment, $A_j$, will influence both the reported treatment, $A_j^{**}$, and the outcome, $Y$. Since $A_j$ is not necessarily measured, this induces confounding in the model. 

The aforementioned shortcomings of ITT analyses in general, the desire to make use of reported treatment indicators, and direct scientific interest each motivate the need for estimators that can estimate the true, optimal DTR, in the presence of nonadherence. To this end, we propose a modified version of G-estimation that directly targets the true optimal DTR. This procedure makes use of modelling nonadherence in the sample and as a result, clinicians and policymakers can integrate both the true efficacy of treatment, and the likelihood of adherence, to best inform recommended treatments for each individual.

\section{G-Estimation with Nonadherence}
\subsection{Modified G-Estimation}
To introduce the modified G-estimation procedure, we first assume that $A_j^*$ are observed for each individual $i$. Let $H_j^*$ denote the history vector with $A_j^*$ recorded in place of $A_j$. For $j=1,\dots,K$, let $\pi_j^*(H_j^*, A_j^*) \triangleq \pr(A_j = 1 \mid H_j^*, A_j^*)$, $\nu_j^*(H_j^*) \triangleq E[\nu_j(H_j)\mid H_j^*, A_j^*]$, and $C_j^*(H_j^*) \triangleq E[C_j(H_j)\mid H_j^*, A_j^*, A_j = 1]$. Define $\widetilde{V}_{K+1} = Y$, and for $j = 1,\dots,K$, $\widetilde{V}_{j} = \widetilde{V}_{j+1} + [A_{j}^\text{opt} - \pi_{j}^*(H_{j}^*)]C_{j}^*(H_{j}^*)$. Finally, take $U_j^*$ to be the set of functions, \begin{multline}
    U_j^*(\psi_j) \triangleq \sum_{i=1}^n \lambda_j^*(H_{i,j}^*)\left\{A_{i,j}^* - \pr(A_{i,j}^* = 1\mid H_{i,j}^*)\right\} \\
    \times\left\{\widetilde{V}_{i,j+1}-\pi_{j}^*(H_{i,j}^*,A_{i,j}^*)C_j^*(H_{i,j}^*; \psi_j) + \theta_j^*(H_{i,j}^*)\right\}. \label{eq::estimating_eq_general}
\end{multline}
These are analogous to equation~\eqref{eq::g_estimation_j} from standard G-estimation. Estimators, $\widehat{\psi}^*_j$, are derived by solving $U_j^*(\psi_j) = 0$. These estimators are consistent, assuming conditions~\ref{cond::IA1} and \ref{cond::IA2}.

\begin{condition}\label{cond::IA1}
    For all $j=1,\dots,K$, \[E[V_{j+1}(H_j)\mid H_j,A_j,A_j^*,A_{j-1}^*,\dots,A_1^*] = E[V_{j+1}(H_j)\mid H_j,A_j].\]
\end{condition}

This condition requires that there is no predictive information contained in the treatment assignment, when a patient's history and true $j$th treatment are known. This can be viewed as a strengthening of the SUTVA. If a patient receives $A_j = 1$ it should not matter whether they were prescribed $A_j^* = 1$ or $A_j^* = 0$. If it does there are two forms of $A_j = 1$.

\begin{condition}\label{cond::IA2}
 For all $j=1,\dots,K$, $E[\nu_j(H_j)\mid H_j^*, A_j^*] = E[\nu_j(H_j)\mid H_j^*]$. 
\end{condition}

This condition requires that the treatment-free model is not predicted by treatment assignment, given a patient's history. Assuming complete adherence, $\nu_j(H_j)$ is functionally independent of $A_j$ and so it is reasonable to assume that it is independent of treatment assignment as well. Condition~\ref{cond::IA2} may be violated if, for instance, past adherence status informs the current treatment, but is not recorded. When factors informing treatment are not recorded there are likely violations of the no unmeasured confounders assumption, questioning the validity of any causal analysis.

The quantity $C_j^*(H_j^*)$ is not a particularly natural quantity to model as it corresponds to the expected contrast given the observable variates and the true treatment being $A_j=1$. Instead, it may be useful to make the following additional independence assumption.

\begin{condition}\label{cond::IA3}
For all $j=1,\dots,K$, $E[C_j(H_j)\mid A_j=1, H_j^*, A_j^*] = E[C_j(H_j)\mid H_j^*, A_j^*]$.
\end{condition}

This condition states that $C_j^*$ is modelled directly from the observable quantities. It requires that there is no mean difference in the contrast between those who actually take the treatment at time $j$ and those who do not, given the observed history and treatment assignments. If previous compliance is related to current compliance then this may be violated. We proceed assuming that either this condition holds or that $C_j^*$ can be specified directly.

\begin{theorem}\label{thm::modified_G_consistency}
    Suppose that conditions~\ref{cond::IA1} and \ref{cond::IA2} hold. Further, suppose that for $j=1,\dots,K$, and for all $n$ individuals, both $\pr(A_{j}^*=1\mid H_{j}^*)$ and $\pi_j^*(H_{j}^*,A_{j}^*)$ are known. If the form of $C_j^*(H_{i,j}^*; \psi_j)$ is correctly specified, then the estimator for $\psi_j$ arising by solving $U_j^*(\psi_j) = 0$, with $U_j^*$ taken from equation~\eqref{eq::estimating_eq_general}, is consistent for the true $\psi_j$. 
\end{theorem}

In practice, $\pr(A_{j}^*=1\mid H_{j}^*)$ and $\pi_j^*(H_{j}^*, A_{j}^*)$ will often not be known. As with standard G-estimation, we can instead specify parametric models for these quantities and jointly estimate all the parameters, while maintaining the established consistency results. If we specify a parametric model for $\theta_j^*(H_j^*)$ to estimate $-\nu_j^*(H_j^*)$, then the modified G-estimation procedure is doubly robust, analogous to the standard case. Supposing that $\pi_j^*(H_j^*, A_j^*)$ and $C_j^*(H_{i,j}^*; \psi_j)$ are correctly specified, if either $\pr(A_j^* = 1\mid H_j^*)$ or $\theta_j^*(H_j^*)$ are correctly specified, then the resulting estimator for $\psi_j$ is consistent. 

The specification of $\lambda_j^*(H_j)$ can be arbitrary as long as it has the same dimension as $\psi_j$. In the class of estimating equations characterized by arbitrary $\lambda_j^*(H_j^*)$, denoting $U_j^*(\psi_j) = \sum_{i=1}^n\lambda_j^*(H_{i,j}^*)\widetilde{U}_{i,j}$, then \citet{Morton_1981} demonstrate that the optimal choice for $\lambda_j^*(H_j^*)$ is \[\lambda_j^*(H_j^*) = E\left(\frac{\partial}{\partial\psi_j}\widetilde{U}_j\mid H_{j}^*\right)E\left(\widetilde{U}_{j}^2\mid H_j^*\right)^{-1}.\] The corresponding quantity for $\lambda_j(H_j)$ was derived by \citet{Robins2004} to achieve locally efficient estimation under complete adherence. While this can be derived in certain settings, the second term in the product is generally quite complex. As a result, we propose the same simplification used for standard G-estimation, taking $\lambda_j^*(H_j^*) = (\partial/\partial\psi_j)C_j^*(H_j^*;\psi_j)$. This will be optimal assuming that all models are correctly specified, that $\text{var}(\widetilde{V}_j\mid H_j^*,A_j^*) = \text{var}(\widetilde{V}_j\mid H_j^*)$, and that both $\text{var}(\widetilde{V}_j\mid H_j^*)$ and $\text{var}(A_j^*\mid H_j^*)$ are constant. 

\subsection{Modeling Nonadherence}
The modified G-estimation procedure relies on being able to accurately model adherence rates in the population. These probabilities are required explicitly as $\pi_j^*(H_j^*, A_j^*)$, and in the model for $C_j^*(H_j^*)$ whenever past treatment is a predictor. The implementation of modified G-estimation thus requires accurate modeling of these terms. If the probabilities cannot be estimated from data and are not known explicitly, the modified G-estimation procedure can proceed using a posited model for patient adherence based on subject-matter expertise. This allows for a sensitivity analysis to be performed quantifying the impact of nonadherence.

If the adherence models are to be estimated from data, then auxiliary information is required. The most straightforward setting occurs when a validation sample is obtained. In this case, at each stage, $j=1,\dots,K$, there is a subset of individuals $i=1,\dots,n_j'$ with both $A_{i,j}^*$ and $A_{i,j}$ measured. Then standard modelling techniques, such as likelihood methods or generalized linear models, can be used to estimate the required probabilities. The same modelling procedures can be pursued if the validation data come from a representative, external sample. Once specified, the estimating equations used to estimate the parameters can be stacked with the estimating equations for the contrast function parameters. 

It is theoretically feasible, using the techniques of \citet{BuonBook} or \citet{Walter_1988}, to estimate the corresponding probabilities from repeated measurements of the prescribed treatment indicators. In practice it is unclear if there is utility in this approach for this situation. In order to apply these techniques we would require multiple, conditionally independent treatment indicators. This strategy is useful when, for instance, the measured binary indicator is disease status that is the result of a possibly faulty test. In our case the binary response represents a treatment prescription without an obvious analogue for how repeated measurements could be taken. We present this possibility in the event that a particular DTR application is well-suited to its application, but do not otherwise pursue it further.

If no auxiliary data are available, modified G-estimation can use existing estimates from external literature. These estimates can be used as though they were the truth, making adjustments to the estimated standard errors. This setting can be viewed as a special case of using external validation data, and as a result is subject to the same asymptotic distribution. 

When no auxiliary data or existing parameter estimates are available, we can conduct a sensitivity analysis. We specify a plausible model for $\pi_j^*(H_j^*, A_j^*)$. Instead of estimating the parameter values, they are specified as though these were the truth. Several sets of plausible parameter values are used in these models, and the modified G-estimation procedure is performed for each of them. The resulting set of estimates can then be taken to indicate the impact of nonadherence on, and provide insight into, the optimal DTR. 

\subsection{Asymptotic Distribution and Inference}
To estimate the stage $j$ contrast function parameters in full generality, we require the joint estimation of parameters arising across five separate sets of estimating equations. Suppose that $\beta_j$ parameterizes $\theta_j^*$ and is estimated by solving $U_{\text{Treatment Free},j}(\beta_j) = 0$, $\alpha_j$ parameterizes $\pi_j^*$ and is estimated by solving $U_{\text{Treatment,j}}(\alpha_j) = 0$, $\gamma_j$ parameterizes $\pr(A_k=1\mid H_j^*)$ and is estimated by solving $U_{\text{Treatment Assignment, j}}(\gamma_j) = 0$, and $\zeta_j$ parameterizes the modified patient history where occurrences of $A_j$ are replaced by their expectations, and is estimated by solving $U_{\text{Modified History,j}}(\zeta_j) = 0$. With $\psi_j$ as the contrast function parameters and $\Theta_j = (\beta_j,\alpha_j,\gamma_j,\zeta_j,\psi_j)$, the modified G-estimation procedure proceeds by solving $U_j^*(\Theta_j) = 0$ where $U_j^*(\Theta_j)$ is the vector formed by concatenating $U_{\text{Modified History, j}}(\zeta_j)$, $U_{\text{Treatment Free, j}}(\beta_j)$, $    U_{\text{Treatment, j}}(\alpha_j)$, $U_{\text{Treatment Assignment, j}}(\gamma_j)$, and $U_{\text{G-Estimation, j}}(\beta_j, \alpha_j, \gamma_j, \zeta_j, \psi_j)$.

For stages $j < K$, $U_j^*(\Theta_j)$ further relies on parameters estimated at future stages, $j' = j+1,\dots,K$. Taking the full set of parameters to be $\Theta = (\Theta_1,\dots,\Theta_K)$, then the modified G-estimation procedure can be framed as solving $U^*(\Theta) = 0$, where \begin{equation} 
U^*(\Theta) = \begin{bmatrix}
    U_K^*(\Theta_K)^\intercal &
    U_{K-1}^*(\Theta_{K-1})^\intercal & \cdots & 
    U_{1}^*(\Theta_1)^\intercal
\end{bmatrix}^\intercal.\label{eq::modified_g_est_full_stack}\end{equation} If any of these parameters are known, or have been previously estimated, the corresponding terms are in the estimating equation are replaced by the known parameter values. This estimation procedure exhibits joint asymptotic normality under the assumption that the data are generated via a \emph{non-exceptional law} \citep{Robins2004}. Briefly, exceptional laws are laws where $\pr(C_j(H_j) = 0) > 0$. In these settings standard asymptotic theory cannot be applied owing to discontinuities in the function for optimal treatment assignment.

\begin{theorem}\label{thm::modified_g_distribution}
    Suppose that $\Theta$ is the solution to $E[U^*(\Theta)] = 0$ and $\widehat{\Theta}^*$ the solution to $\frac{1}{n}\sum_{i=1}^n U^*(\Theta; H_{i,K}^*, A_{i,K}^*) = 0$, where $U^*$ is established in equation~(\ref{eq::modified_g_est_full_stack}). Taking $\Psi = (\psi_1,\dots, \psi_K)$ to be the contrast function parameters contained in $\Theta$ and $\widehat{\Psi}^* = (\widehat{\psi}_1^*,\dots,\widehat{\psi}_K^*)$ to be the estimated contrast function parameters from $\widehat{\Theta}^*$, then under the regularity conditions set out by \citet{Robins2004}, as $n\to\infty$, $\sqrt{n}\left(\widehat{\Psi}^*-\Psi\right) \stackrel{d}{\longto} N\left(\mathbf{0}, \Sigma_\Psi\right)$. Here $\Sigma_\Psi = I_{\Psi}\Sigma_{\Theta}I_{\Psi}$, $I_{\Psi}$ is the diagonal matrix with $1$s on the diagonal entries corresponding to the locations of the $\Psi$ parameters in $\Theta$, and $\Sigma_\Theta$ is a sandwich variance matrix based on $U^*$, \[\Sigma_\Theta = E\left[-\frac{\partial}{\partial\Theta}U^*(\Theta)\right]^{-1}E\left[U^*(\Theta)U^{*}(\Theta)^\intercal\right]\left(E\left[-\frac{\partial}{\partial\Theta}U^*(\Theta)\right]^{-1}\right)^\intercal.\] Under the assumptions of Theorem~\ref{thm::modified_G_consistency}, $\Psi$ corresponds to the true contrast function parameters.
\end{theorem}

Asymptotic normality follows from standard M-estimation theory under the previously described assumption of non-exceptional laws \citep{Robins2004}. The consistency established in Theorem~\ref{thm::modified_G_consistency} does not require the assumption of a non-exceptional law. Techniques to address exceptional laws have been studied and we expect that the resulting recommendations will apply for the modified G-estimation procedure \citep{Chakraborty_2009,Moodie_2010,mn_q_learning}. 

\subsection{Prescribed, Actual, and Reported Treatments}
The modified G-estimation procedure was presented assuming that the prescribed treatments, $A_j^*$, are used to make inference regarding the true treatment, $A_j$. If instead we wish to use the reported treatment, $A_j^{**}$, modified G-estimation proceeds as described substituting $A_j^{**}$ for $A_j^*$ and defining the analogous quantities. The primary difficulty in doing so stems from the need to specify the model $\pi_j^{**}(H_j^{**}, A_j^{**}) = \pr(A_j=1\mid A_j^{**},H_j^{**})$. This model is well-defined statistically, however, it is specified using a reverse causal structure. Specifically, $A_j$ is an antecedent of $A_j^{**}$, rather than the reverse. As a result, it may be more challenging to have a well-informed, subject-matter perspective on the terms in the model. 

If the correct model is specified for these quantities, and the remaining models are updated to use $A_j^{**}$ in place of $A_j^*$, then both Theorems~\ref{thm::modified_G_consistency} and \ref{thm::modified_g_distribution} will hold. This is an important result as, to reiterate our previous discussion, replacing $A_j$ with $A_j^{**}$ produces invalid causal estimates in general. This problem is circumvented with the modified G-estimation procedure.

\subsection{Pseudo Outcomes and Optimal Treatments}\label{sec::nonadherence_pseudo_outcomes}
Theorems~\ref{thm::modified_G_consistency} and \ref{thm::modified_g_distribution} rely on our ability to accurately construct the pseudo outcomes, defined as $\widetilde{V}_{j} = \widetilde{V}_{j+1} + [A_{j}^\text{opt} - \pi_j^*(H_{j}^*)]C_j^*(H_{j}^*)$. In the event of errors in the individual level factors, $X_j$, it has been shown that accurate construction of the pseudo outcomes represents a substantial hurdle to exactly consistent estimation of the optimal DTR \citep{Spicker2019_ME}. Even when the indexing parameters for the treatment propensity model and the contrast function are known exactly, measurement error can cause the optimal indicator, $A_j^\text{opt}$, to be incorrect. Despite this, use of the approximately correct pseudo outcomes formed by replacing $A_j^\text{opt}$ with its estimated value resulted in acceptable performance in simulations.

Nonadherence introduces a similar concern. If both $\pi_j^*(H_j^*)$ and $C_j^*(H_j^*)$ are correctly specified, and the corresponding estimators for $\alpha_j$ and $\psi_j$ are consistent, this will not guarantee consistency of the plug-in estimator for $A_j^\text{opt}$. If there are no past treatment indicators in the contrast function then $C_j^*(H_j^*) = C_j(H_j)$ and consistent estimation of $\psi_j$ implies consistent estimation of $A_j^\text{opt}$. If any $A_\ell$, for $\ell < j$, are predictors in $C_j(H_j)$, then these indicators will be replaced by $\pr(A_\ell = 1\mid H_j^*)$ in the model for $C_j^*(H_j^*)$. The concern regarding pseudo outcome construction is with the addition of the term $A_j^\text{opt}C_j^*(H_j^*)$. The same asymptotic results are obtained by replacing the term $A_j^\text{opt}C_j^*(H_j^*)$ with any term with expectation $E[A_j^\text{opt}C_j^*(H_j^*)\mid H_j^*,A_j^*]$.

For exposition, suppose that the only treatment indicator in $C_j(H_j)$ is $A_{j-1}$. Then, \begin{multline} E[A_{j}^\text{opt}C_j(H_j)\mid H_j^*, A_j^*] = \pi_{j-1}^*(H_{j}^*)I\left\{C_{j}(H_{j}^*, A_{j-1}=1)>0\right\}C_{j}(H_{j}^*,A_{j-1}=1) \\
+\left[1-\pi_{j-1}^*(H_{j}^*)\right]I\left\{C_{j}(H_{j}^*, A_{j-1}=0)>0\right\}C_{j}(H_{j}^*,A_{j-1}=0).\label{eq::expected_po}\end{multline}

Based on the required modelling this term is computable. This computation could be used to form a pseudo outcome that is valid when treatment indicators are included as predictors in the contrast function. Taking $\Delta_j = E[A_{j}^\text{opt}C_j(H_j)\mid H_j^*, A_j^*]$ as in equation~\eqref{eq::expected_po}, then $\widetilde{V}_j = \widetilde{V}_{j+1} + \Delta_j$ is a valid pseudo outcome. In practice, the contrast function and its parameters will not be precisely known. Similarly, the adherence probabilities will be estimated, either from the current sample or from past data. As such, the theoretical benefit obtained by deriving exact expressions for the pseudo outcomes may be undermined by errors in the modelling procedures. We will continue to use the approximately correct pseudo outcomes, that are exactly correct whenever past treatment is not used to tailor decisions, and note that consistent estimators for the pseudo outcomes are available when needed.

\subsection{Multiple Treatment Alternatives}\label{sec::multiple_treatment_alternatives}
The proposed procedures have been presented with binary treatments. While this assumption is common in the DTR literature, these techniques can be extended to arbitrary categorical treatments. It is conceptually straightforward to accommodate treatment $A_j \in \mathcal{A}_j \neq \{0,1\}$. The primary substitution required is to take \[Q_j(H_j, A_j) = \nu_j(H_j) + \sum_{a_j \in \mathcal{A}_j} I(A_j = a_j)C_{j,a_j}(H_j;\psi_{j,a_j}).\] Here, $C_{j,a_j}$ is the contrast function at stage $j$, corresponding to treatment $a_j \in \mathcal{A}_j$, given by $C_{j,a_j}(H_j) = E[\widetilde{V}_{j+1}\mid H_j, A_j = a_{j}] - E[\widetilde{V}_{j+1}\mid H_j, A_j = 0]$. With nonadherence, these quantities extend in the same way as in the binary case. The nonadherence rates require a model for each actual treatment, $\pi_{j,a_j}^*(H_j^*,A_j^*) \triangleq\pr(A_j = a_j\mid H_j^*,A_j^*)$.

While it is common to assume binary treatments in the DTR literature, this practice is worth considering in more depth when addressing nonadherence. Implicit in our discussions is the assumption that a patient prescribed treatment $A_j^* = 1$, who is nonadherent, takes the same treatment as a patient who is adherent to $A_j^* = 0$. Suppose that $A_j = 1$ is an experimental treatment contrasted with $A_j = 0$ for standard care. It is likely that a patient who is nonadherent to $A_j^* = 1$ will be taking a third treatment, say $A_j = -1$, corresponding to no active treatment. The alternative scenario, where $A_j^*=0$ is seemingly more likely to introduce a third category under the assumption of nonadherence. When treatment refers to a specific pharmacological intervention it is unlikely that a patient who is nonadherent to $A_j^* = 0$ will gain access to the active treatment under study. However, when treatment refers to, for instance, a lifestyle intervention such as weight loss or exercise, it is plausible that nonadherent individuals will switch from one treatment category to the other. There is not a universal assumption regarding the adherence mechanisms. Instead, the precise categorization will depend on the specific subject matter. 

Consider the MACS analysis presented in Section~\ref{sec::MACS}. In our analysis, treatment refers to the decision to start Zidovudine (AZT) therapy. An individual who starts therapy at stage $j$ will have $A_j = 1$. Once prescribed, the patient remains on AZT indefinitely, and so the analysis considers only the timing of the therapy \citep{Wallace_2016}. We know that some patients are nonadherent to treatment, however, nonadherence to $A_j^* = 1$ is unlikely to result in a patient who has taken no AZT. Instead, nonadherence to $A_j^* = 1$ often corresponds to partial adherence to AZT therapy. An analysis of MACS may be better served by considering three treatment categories: $A_j = 1$, full adherence, $A_j = 0$, nonadherence, and $A_j = -1$, partial adherence. These considerations increase the modelling complexity, however, the modified G-estimation procedure can accommodate these more complex situations where required.

\section{Simulation Studies}
\subsection{Misclassification Dependent on Tailoring Variates}
\begin{figure}
    \centering
    \includegraphics[width=\textwidth]{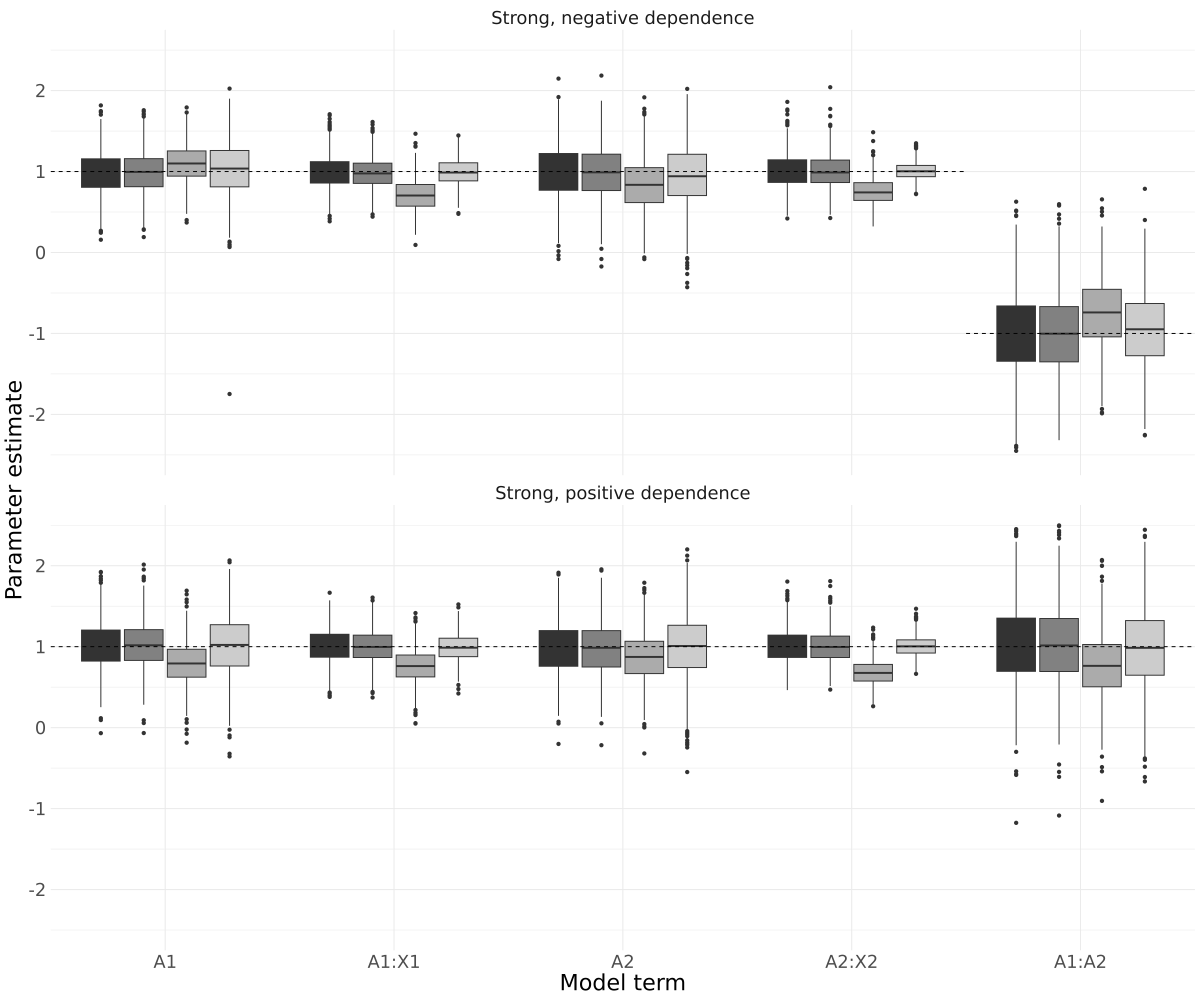}
    \caption{Estimated parameter values across $1000$ simulation runs for a two-stage dynamic treatment regime comparing $\psi_{22} = -1$ (strong, negative dependence) and $\psi_{22} = 1$ (strong, positive dependence). For each model term, from left to right, we compare the results of the modified G-estimation with known adherence rates (darkest), the modified G-estimation with estimated adherence rates (moderately dark), the standard G-estimation using prescribed treatment (moderately light), and the standard G-estimation using the truth (lightest).}
    \label{fig::simulation_1}
\end{figure}

First, we consider a two-stage DTR with two independent tailoring variates $X_1 \sim N(1,1)$ and $X_2 \sim N(1,4)$. We take $\pr(A_j^* = 1 \mid X_j) = \expit(X_j)$, for $j=1,2$, where $\expit(x) = (1+\exp(-x))^{-1}$ is the inverse logistic function. Given assigned treatment and $X_j$, we specify $\pr(A_j = 1\mid A_j^*, X_j) = \expit(-4.6 - 0.83X_j + 7.5A_j^*)$. The parameter values result in fairly low misclassification rates with values of approximately $0.01$ and $0.05$ for those prescribed $A_j^* = 0$ and $A_j^* = 1$, respectively. At stage one the contrast function is $1 + X_1$ and at stage two it is $1 + X_2 + \psi_{22}A_1$, where $\psi_{22}$ is varied across the grid $\{-1,-0.1,0,0.1,1\}$. The treatment-free component is $X_1$, and the outcome follows a normal distribution with variance $2$. Data are simulated with a sample size of $1000$, using an internal validation sample of $30\%$. The simulations are repeated $1000$ times, and box plots of the estimated contrast function parameters for $\psi_{22} \in \{-1,1\}$ are shown in Fig.~\ref{fig::simulation_1}, and the remaining plots are available in the web supplementary material. Estimation compares an as-treated analysis using standard G-estimation where the true treatment variables are available, standard G-estimation using the prescribed treatment in place of truth, modified G-estimation with estimated nonadherence probabilities, and modified G-estimation assuming the nonadherence rates were known.

The naive analysis produces results that are, predictably, biased for the true parameters. Both of the corrections, and the as-treated analysis perform similarly with approximately unbiased parameter estimates. In some settings there is an evident reduction in the variation of the estimates when using the modified G-estimation procedure, rather than that based on the truth. The patterns observed from these two scenarios are visible across the scenarios with $\psi_{22}\in\{-0.1,0,0.1\}$ shown in the web supplementary material.

\subsection{Validation Set Sizing}
\begin{figure}
    \centering
    \includegraphics[width=\textwidth]{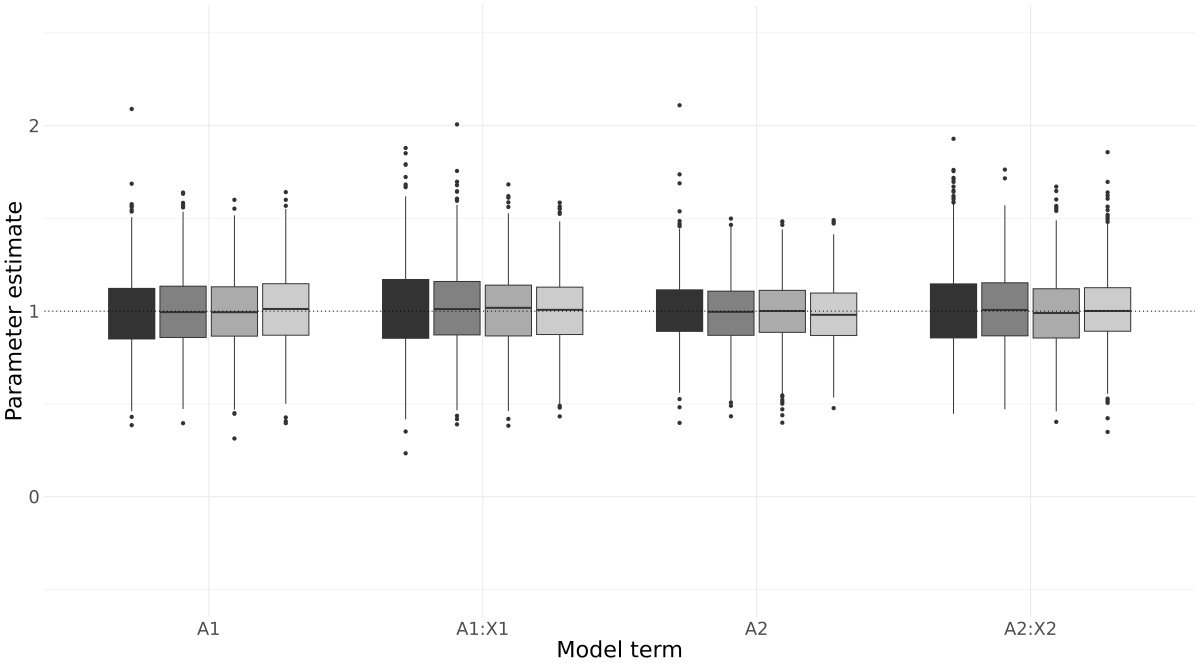}
    \caption{Estimated parameter values across $1000$ simulation runs for a two-stage dynamic treatment regime varying the size of the validation sample. For each model term the box plots from left to right correspond to validation samples of size $10\%$ (darkest), $20\%$ (moderately dark), $30\%$ (moderately light), and $50\%$ (lightest). The sample size is $1000$, and the dotted line indicates the true parameter value.}
    \label{fig::simulation_2}
\end{figure}

The second simulation considers the same data generating process as the first experiment, with $\psi_{22}=0$. We vary the size of the replication set, considering $10\%$, $20\%$, $30\%$, and $50\%$ internal validation samples, with a sample size of $1000$. All models were fit for each of the estimation techniques described for simulation one. The standard G-estimation procedures and the modified G-estimation procedure with known adherence rates are independent of the size of the validation sample, and so Fig.~\ref{fig::simulation_2} reports the box plots for the estimated parameter values across the modified G-estimation procedure with estimated adherence rates. 

Increasing the size of the validation sample from $10\%$ to $20\%$ shows a notable improvement in the variation of the estimators. Beyond this point the method appears to perform comparably well in all settings, with only minor changes in the estimator variance.

\subsection{Asymptotic Coverage Probabilities}
\begin{figure}
    \centering
    \includegraphics[width=\textwidth]{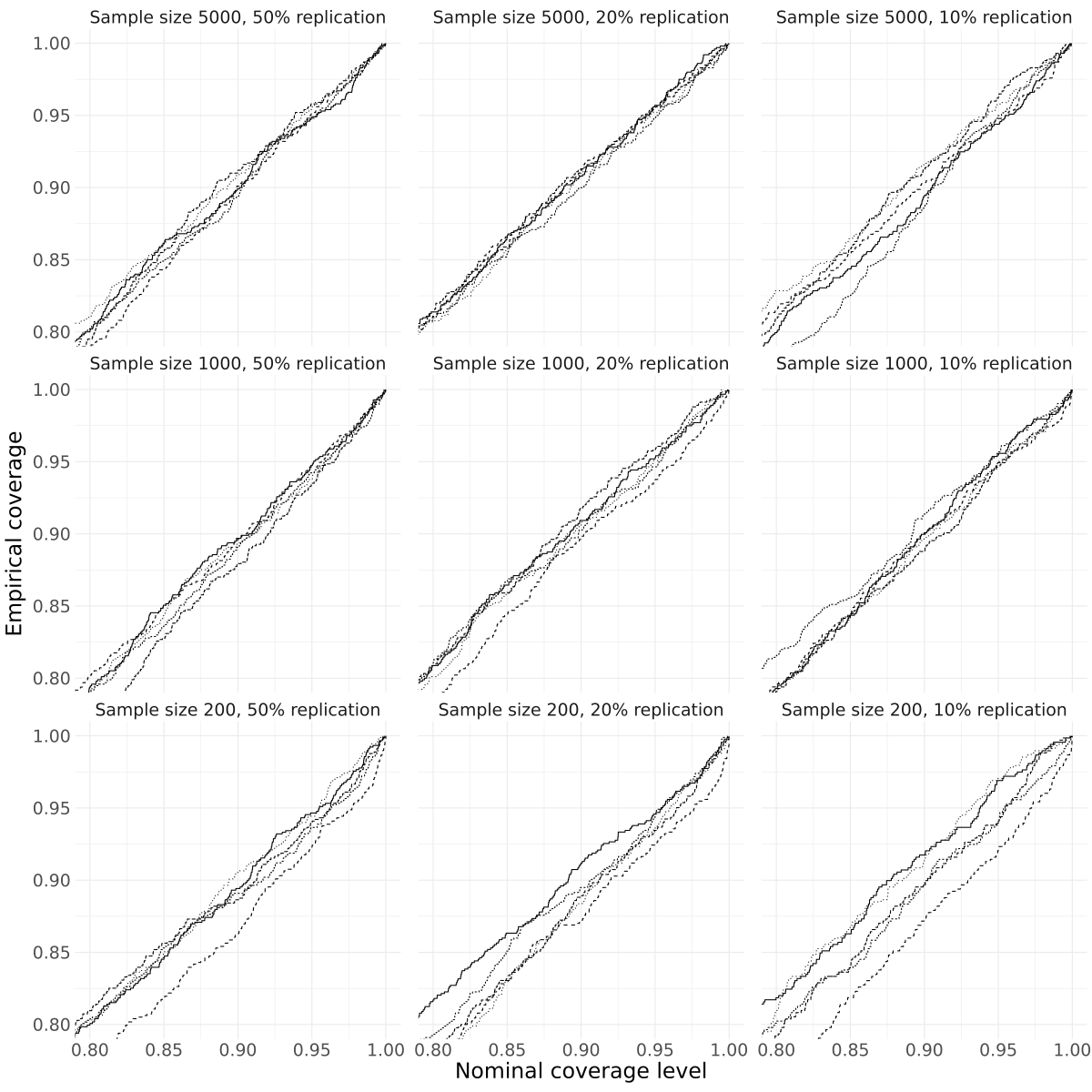}
    \caption{Empirical and novel coverage probability based on $1000$ simulation runs for the estimated parameter values over varying sample sizes ($n$) and validation proportions ($p$). Lines are shown for parameters $\psi_{10}$ (solid), $\psi_{11}$ (short dash), $\psi_{20}$ (dash-dash-dot), $\psi_{21}$ (dash), and $\psi_{22}$ (dot).}
    \label{fig::simulation_3}
\end{figure}

The third simulation investigates the coverage probabilities obtained via the use of sandwich estimation techniques. The same tailoring variables and adherence rates are used as in the first two scenarios. We take $\pr(A_1^* = 1\mid X_1) = \expit(0.5 + X_1)$ and $\pr(A_2^* = 1 \mid X_2) = \expit(-0.5 + X_2)$. The stage $j$ contrast function is $1 + X_j + \psi_{j2}A_j^*$, where $\psi_{12} = 1$ and $\psi_{22} = -1$. The treatment-free model is $X_1 + 0.5A_1^*$. Nine scenarios are considered, selecting the sample size to be low ($n=200$), medium ($n=1000$), or large ($n=5000$) with the validation sample being small ($10\%$), medium ($20\%$), or large ($50\%$). Each scenario is repeated $1000$ times, and the corresponding standard errors are estimated based on approximate sandwich estimation techniques computed via numerical differentiation. Figure~\ref{fig::simulation_3} summarizes the empirical coverage probabilities across the various scenarios and different model parameters. 

The results suggest that coverage is well calibrated, even in small samples with a low validation percentage, though there is a notable improvement with increasing sample size. The estimators for $\psi_{21}$ exhibit the worst coverage behavior, though coverage still tends to be approximately at the nominal levels. The simulated scenario was not under an exceptional law and as such standard asymptotic theory applies.

\subsection{Reported Treatment Correction}
Finally, we use reported treatments rather than prescribed treatments. We consider a two-stage DTR, with $X_1$ and $X_2$ uniformly distributed on $\{-1,0,1\}$, independently. For $j=1,2$, $\pr(A_j = 1\mid X_j) = 0.5 + 0.3X_j$, and $\pr(A_j^{**} = 1\mid A_j,X_j) = A_j(0.9 - 0.05X_j) + (1-A_j)(0.05 + 0.045X_j + 0.005X_j^2)$. The stage one contrast function is $1 + X_1$, and the stage two is $1 + \psi_{21}A_1$, where $\psi_{21}$ is selected across $\{-1,-0.1,0,0.1,1\}$. Data is generated with a sample size of $1000$, using $30\%$ internal validation samples. We compare the modified G-estimation procedure with estimated adherence rates, the naive procedure based on reported treatment, and an analysis using standard G-estimation assuming the truth were available. Table~\ref{tb::MSE_reported_trt} contains the estimated MSEs.

\begin{table}[ht]
    \centering
    \caption{$100$ times the observed MSE, based on $1000$ replicated simulations, for contrast function parameter estimates in a two-stage DTR, where reported treatments are used as a misclassified version of the truth. The results are based on a sample size of $n=1000$, with a validation set of $30\%$, and they compare the corrected estimators, to those that naively apply G-estimation without correction, to those that are obtained when the true treatment status is reported.}
    \begin{tabular}{lcccccccccccc}
      & \multicolumn{4}{c}{Corrected} & \multicolumn{4}{c}{Naive} & \multicolumn{4}{c}{Truth}\\
     & $\psi_{10}$ & $\psi_{11}$ & $\psi_{20}$ & $\psi_{21}$ & $\psi_{10}$ & $\psi_{11}$ & $\psi_{20}$ & $\psi_{21}$& $\psi_{10}$ & $\psi_{11}$ & $\psi_{20}$ & $\psi_{21}$  \\
        $\psi_{22}=-1$   & 3.4 & 3.3 & 2.8 & 6.4 & 7.8  & 30.6 & 11.2 & 15.4 & 2.3 & 2.0 & 2.2 & 4.3 \\
        $\psi_{22}=-0.1$ & 3.6 & 2.8 & 2.7 & 5.5 & 12.5 & 22.5 & 7.2  & 4.2  & 2.0 & 1.9 & 2.1 & 3.9 \\
        $\psi_{22}=0$    & 3.5 & 2.9 & 2.5 & 5.6 & 12.5 & 21.6 & 6.8  & 4.3  & 2.2 & 2.0 & 2.2 & 4.4 \\
        $\psi_{22}=0.1$  & 3.4 & 3.2 & 2.6 & 5.9 & 13.0 & 21.2 & 6.7  & 4.8  & 2.1 & 2.1 & 2.3 & 4.8 \\
        $\psi_{22}=1$    & 3.7 & 2.8 & 2.5 & 6.4 & 15.5 & 13.7 & 3.9  & 15.7 & 2.4 & 1.9 & 2.2 & 4.8 
    \end{tabular}
    \label{tb::MSE_reported_trt}
\end{table}

As expected, the estimators based on truth have the lowest MSE across all scenarios. The modified G-estimation procedure performs comparably, despite the additional modelling requirements. The naive estimators have larger MSEs, making them unreliable as a means of estimating treatment efficacy. Additionally, unlike in the previous scenarios, the naive analysis here cannot be reinterpreted as a causal estimate of a different estimand.

\section{Data Analysis}\label{sec::MACS}
We demonstrate the utility of the proposed corrections by considering an analysis of the MACS data \citep{MACS}. Our analysis follows \citet{Wallace_2016} and \citet{HernanMACS} and considers a restricted set of data to address a comparatively simple question regarding the timing of intervention with a particular antiretroviral drug, AZT. Because AZT became available in March of 1986, our analysis is restricted to those who were HIV positive but AIDS free at this time. We further restrict our sample to include only the first two eligible visits for any individual for clarity of illustration. Individuals were recruited over study waves so the calendar time of the two eligible visits differs between individuals. 

The outcome is the count of a patient's CD4 cells, a type of white blood cell critical for immune response, at the visit immediately following their second eligible visit. We consider the lab results on CD8 cell counts, white blood cell counts, red blood cell counts, platelet counts, both their systolic and diastolic blood pressure, their weight, and an indicator of recent symptoms, that has a value of one if the patient reports any of a fever, oral candidiasis, diarrhea, weight loss, oral hairy leukoplakia, or herpes zoster in their recent medical history. Additionally, we use data from a questionnaire administered in October 1998 to assess patient adherence \citep{MACS_Adherence}. While these data correspond to self-reported adherence data, $A_j^{**}$ in our notation, we treat them as though they were true reports of actual treatment for the individuals who responded. These adherence data only began to be collected in 1998, $12$ years after eligibility into our subset for analysis. It is possible that adherence early in the study was substantially different from the later observed adherence information, however, we assume that adherence is approximately consistent throughout the study. As discussed in Section~\ref{sec::multiple_treatment_alternatives}, most patients who reported being nonadherent remained partially adherent. To simplify our analysis, we will continue to make a binary treatment distinction, though a complete analysis of these data may wish to define multiple treatment alternatives. 

Define $A_j^*$ to represent whether AZT was started during period $j$, $A_j^* = 1$, or not. Individuals prescribed AZT remain on AZT for the remainder of the study. We assume that an individual who has not started on AZT does not take AZT, that is to say $\pr(A_j = 0 \mid A_j^* = 0) = 1$. Denote the year of birth for an individual as $U$, their CD4 count at visit $j$ as $C_j$, their CD8 count at visit $j$ as $K_j$, their red blood cell count at visit $j$ as $R_j$, their white blood cell count at visit $j$ as $W_j$, the platelet count at visit $j$ as $P_j$, their systolic blood pressure at visit $j$ as $T_j$, their diastolic blood pressure at visit $j$ as $D_j$, an indicator representing their AIDS status at visit $j$ as $F_j$, their body weight in pounds at visit $j$ as $B_j$, and the indicator of the presence of symptoms at stage $j$ as $S_j$. 

The nonadherence model is fit using logistic regression on the validation data. We find that the adherence rates appear to be consistent between the first and second stage, and select the model $\text{logit}(\pr(A_j = 1\mid H_j^*, A_j^*=1)) = \alpha_0 + \alpha_1 U + \alpha_2\log(D_j) + \alpha_3\log(T_j) + \alpha_4\log(C_j)$. With the estimated adherence model, we construct an optimal DTR, following \citet{Wallace_2016} for the considered functional forms. We conduct a complete case analysis. Models are fit using information from $2850$ patients with information collected across a total of $8550$ visits. The adherence model is fit using $766$ responses from a total of $220$ patients. 

The stage one outcome model is $\beta_{10} + \beta_{11}C_1 + \beta_{12}\log C_1 + A_1\times(\psi_{11} + \psi_{12}U + \psi_{13}\log C_1 + \psi_{14}S_1 )$. For the second stage, the outcome model is $\beta_{20} + \beta_{21}C_1 + \beta_{22}\log C_1 + \beta_{23}C_2 + \beta_{24}\log C_2 + A_2\left(\psi_{21} + \psi_{22}U + \psi_{23}\log C_2 + \psi_{24}S_2 \right)$. For $j=1,2$ we specify the corresponding treatment prescription models as $\text{logit}(\pr(A_1^*=1\mid \cdot)) = \gamma_{j1} + \gamma_{j2}C_j + \gamma_{j3}K_j + \gamma_{j4}R_j + \gamma_{j5}W_j + \gamma_{j6}P_j$. To standardize the magnitude of coefficients, we transform $U$ to represent the patient's age in 1986, rather than their birth year. Table~\ref{tb::parameter_estimates_MACS} displays the estimates for both a naive analysis and one using the modified G-estimation procedure. Alongside the point estimates, $95\%$ confidence intervals using $1000$ bootstrap replicates are presented.

\begin{table}[ht]
\centering
\caption{Estimated contrast function parameters, with 95\% bootstrapped confidence intervals, based on a naive analysis of MACS assuming full adherence and an analysis based on the modified G-estimation procedure accounting for nonadherence.}
\begin{tabular}{rrrrrrr}
    & \multicolumn{3}{c}{Naive} & \multicolumn{3}{c}{Modified G-estimation} \\
    & Lower & Estimate & Upper & Lower & Estimate & Upper \\
    $\psi_{11}$ & -4647.81 & 8254.73 & 20884.43 & -25434.32 & 40623.92 & 100192.05 \\
    $\psi_{12}$ & -10.13 & -3.93 & 2.50 & -48.20 & -19.51 & 12.76 \\
    $\psi_{13}$ & -137.68 & -49.61 & 29.90 & -695.41 & -182.41 & 171.44 \\
    $\psi_{14}$ & -198.44 & -82.76 & 46.39 & -709.97 & -193.30 & 284.67 \\ 
    $\psi_{21}$ & -1065.59 & 5528.36 & 12594.28 & 3037.40 & 35922.73 & 66075.74 \\
    $\psi_{22}$ & -6.14 & -2.66 & 0.64 & -32.62 & -17.53 & -1.34 \\
    $\psi_{23}$ & -74.90 & -34.41 & 8.85 & -364.82 & -120.10 & 174.50 \\
    $\psi_{24}$ & -124.93 & -70.34 & -13.56 & -545.72 & -238.67 & 8.81 \\
\end{tabular}
\label{tb::parameter_estimates_MACS}
\end{table}

Both techniques have highly variable estimates. Of note are the results for $\psi_{22}$ and $\psi_{24}$. These factors, that correspond to the tailoring effects of age and the presence of symptoms, respectively, are the only two factors that are found to be statistically significant at a $5\%$ significance level. The corrected estimation technique finds that $\psi_{22}$ is significantly different from $0$, with a negative impact on the outcome, while the naive analysis misses this factor. Conversely, the naive analysis indicates significance of $\psi_{24}$, also with a negative impact on the outcome, while the modified procedure does not. 

Beyond the estimated coefficients we consider the agreement of treatment assignment across the bootstrap replicates between the two methods. The median proportion of agreement between the methods at stages one and two were $0.913$ and $0.962$, respectively. While this is a fairly high proportion of agreement it is worth quantifying the level of nonadherence in these data. Approximately $90\%$ of individuals who were assigned AZT were fully adherent. Roughly $2\%$ of respondents had $A_1^* = 1$ and roughly $5\%$ of respondents had $A_2^* = 1$. We have assumed that those with $A_j^* = 0$ are adherent, and so this corresponds to an approximately $0.5\%$ rate of nonadherence across the study. This small degree of nonadherence shifts the optimal treatment assignment for between $3.8\%$ and $8.7\%$ of individuals. While the shortcomings of our analysis render it unlikely that our specific effect estimates are indicative of the underlying reality, this analysis makes clear the concerns with ignoring adherence information. Even small amounts of nonadherence can greatly impact the estimation. 

\section{Discussion}
Nonadherence is a pervasive concern in medical data that can, in many settings, invalidate causal analyses performed on such data. While the effects of nonadherence are well understood broadly, to date there were no existing techniques for optimal DTR estimation which corrected for the impacts of nonadherence. The proposed modified G-estimation restores the desirable properties of G-estimation for optimal DTR estimation when data are subject to nonadherence. This method provides practitioners with a corrective procedure when sufficient auxiliary information is available to correctly model the rates of nonadherence in the study population. When such data is not available, the technique can be applied to determine how sensitive the estimated DTR is to hypothesized degrees of nonadherence.

It is important to point out that the proposed techniques rely on correctly specified models for the nonadherence mechanisms in order to remain consistent, and on various conditions regarding the dependence structures in the data. Moreover, our methods did not explore the explicit use of multiple error-prone treatment indicators, say both the prescribed and reported treatments. Both of these areas remain important lines of inquiry for future work.






\bibliographystyle{unsrtnat}
\bibliography{references.bib}

\pagebreak 

\appendix

\section{Proofs of Results}

\begin{proof}[Theorem~\ref{thm::modified_G_consistency}]
    First, we show that \[E[\widetilde{V}_{j+1}\mid H_j^*,A_j^*] = \nu_j^*(H_j^*) + \pi_j^*(H_j^*, A_j^*)C_j^*(H_j^*).\] Then we show that $E[U_j^*(\psi_j)] = 0$. We begin using induction. First, for $j=K+1$, we have $\widetilde{V}_j = Y$ and so \begin{align*}
        &E[Y\mid H_K^*, A_K^*] \\
        &= E\left\{E\left[Y\mid H_K,A_K,\overline{A}_K^*\right]\mid H_K^*, A_K^*\right\} \\
        &= E\left\{E\left[Y\mid H_K,A_K\right]\mid H_K^*, A_K^*\right\} &&\text{Condition~\ref{cond::IA1}} \\
        &= E\left\{Q_K(H_K,A_K)\mid H_K^*, A_K^*\right\} \\
        &= E\left\{\nu_K(H_K) + A_KC_K(H_K;\psi_K)\mid H_K^*, A_K^*\right\} \\
        &= E\left\{\nu_K(H_K)\mid H_K^*, A_K^*\right\} + E\left\{A_KC_K(H_K;\psi_K)\mid H_K^*, A_K^*\right\} \\
        &= \nu_K^*(H_K^*) + \pr(A_K=1\mid H_K^*,A_K^*)E\left\{C_K(H_K;\psi_K)\mid A_K=1, H_K^*, A_K^*\right\} \\
        &= \nu_K^*(H_K^*) + \pi_K^*(H_K^*,A_K^*)C_K^*(H_K^*) &&\text{Condition~\ref{cond::IA3}}.
    \end{align*} Next, suppose take the inductive hypothesis (I.H.) to be that this expression holds for $j=K,\dots,k+2$, and consider \begin{align*}
        &E[\widetilde{V}_{k+1}\mid H_{k}^*,A_k^*] \\
        &= E[\widetilde{V}_{k+2} + \left\{A_{k+1}^\text{opt} - \pi_{k+1}^*(H_{k+1}^*,A_{k+1}^*)\right\}C_{k+1}^*(H_{k+1}^*)\mid H_{k}^*,A_k^*] \\
        &= E\left\{E\left[\widetilde{V}_{k+2}\mid H_{k+1}^*,A_{k+1}^*\right] + \left\{A_{k+1}^\text{opt} - \pi_{k+1}^*(H_{k+1}^*,A_{k+1}^*)\right\}C_{k+1}^*(H_{k+1}^*)\mid H_k^*, A_k^*\right\}  \\
        &= E\left\{\left.\nu_{k+1}^*(H_{k+1}^*) + \pi_{k+1}^*(H_{k+1}^*,A_{k+1}^*)C_{k+1}^*(H_{k+1}^*) \right.\right.\\
        &\quad\quad \left.+ \left\{A_{k+1}^\text{opt} - \pi_{k+1}^*(H_{k+1}^*,A_{k+1}^*)\right\}C_{k+1}^*(H_{k+1}^*)\mid H_k^*, A_k^*\right\} &&\text{I.H}\\
        &= E\left\{\nu_{k+1}^*(H_{k+1}^*) + A_{k+1}^\text{opt}C_{k+1}^*(H_{k+1}^*)\mid H_k^*, A_k^*\right\} \\
        &= E\left\{E[\nu_{k+1}(H_{k+1}) + A_{k+1}^\text{opt}C_{k+1}(H_{k+1})\mid H_{k+1}^*,A_{k+1}^*]\mid H_k^*, A_k^*\right\} \\
        &= E\left\{\nu_{k+1}(H_{k+1}) + A_{k+1}^\text{opt}C_{k+1}(H_{k+1})\mid H_k^*, A_k^*\right\} \\
        &= E\left\{V_{k+1}(H_{k+1})\mid H_k^*, A_k^*\right\} \\
        &= E\left\{E\left\{V_{k+1}(H_{k+1})\mid H_k,A_k,\overline{A}_k^*\right\}\mid H_k^*, A_k^*\right\} \\
        &= E\left\{\nu_k(H_k) + A_{k}C_k(H_k)\mid H_k^*, A_k^*\right\} &&\text{Condition~\ref{cond::IA1}}\\
        &= \nu_k^*(H_k^*) + \pr(A_k=1\mid H_k^*,A_k^*)E\left\{C_k(H_k)\mid A_k=1, H_k^*, A_k^*\right\} \\
        &= \nu_k^*(H_k^*) + \pi_k^*(H_k^*,A_k^*)C_k(H_k^*). &&\text{Condition~\ref{cond::IA3}}
    \end{align*} In addition to the independence assumptions and the inductive hypothesis, we also used the fact that $C_k(\cdot)$ is correctly specified. In the event (as will be the case in practice) that we are using the estimated versions instead, all of these equalities hold almost surely (assuming that $\widehat{\psi}_j$ are almost surely consistent for $\psi_j$). With these expected pseudo-outcome (E.P.O) results established, we can show that $E[U_j^*(\psi_j)] = 0$. First, consider the expectation, conditional on $\{H_j^*,A_j^*\}$ \begin{align*}
        &E\left[U_j^*(\psi_j)\mid H_j^*,A_j^*\right] \\
        &= \sum_{i=1}^n\lambda_j^*(H_{i,j}^*)\left\{A_{i,j}^* - \pr(A_{i,j}^* = 1\mid H_{i,j}^*)\right\}\\
        &\quad\quad\times\left\{E[\widetilde{V}_{i,j+1}\mid H_{i,j}^*,A_{i,j}^*]-\pi_{j}^*(H_{i,j}^*,A_{i,j}^*)C_j^*(H_{i,j}^*; \psi_j) + \theta_j^*(H_{i,j}^*)\right\} \\
        &= \sum_{i=1}^n\lambda_j^*(H_{i,j}^*)\left\{A_{i,j}^* - \pr(A_{i,j}^* = 1\mid H_{i,j}^*)\right\}\left\{\nu_j^*(H_{i,j}^*) + \theta_j^*(H_{i,j}^*)\right\} &&\text{E.P.O.}
    \end{align*} Taking this result, we can consider the expectation conditional on just $H_j^*$, which gives \begin{align*}
         &E\left[U_j^*(\psi_j)\mid H_j^*\right] \\
         &= \sum_{i=1}^n\lambda_j^*(H_{i,j}^*)\left\{E[A_{i,j}^*\mid H_{i,j}^*] - \pr(A_{i,j}^* = 1\mid H_{i,j}^*)\right\}\left\{\nu_j^*(H_{i,j}^*) + \theta_j^*(H_{i,j}^*)\right\} &&\text{Condition~\ref{cond::IA2}} \\
         &= \sum_{i=1}^n\lambda_j^*(H_{i,j}^*)\left\{\pr(A_{i,j}^*=1\mid H_{i,j}^*) - \pr(A_{i,j}^* = 1\mid H_{i,j}^*)\right\}\left\{\nu_j^*(H_{i,j}^*) + \theta_j^*(H_{i,j}^*)\right\} \\
         &= 0.
    \end{align*} This will hold so long as the residual term \[E[\widetilde{V}_{i,j+1}\mid H_{i,j}^*,A_{i,j}^*]-\pi_{j}^*(H_{i,j}^*,A_{i,j}^*)C_j^*(H_{i,j}^*; \psi_j) + \theta_j^*(H_{i,j}^*),\] is independent of $H_{i,j}^*$. This is true under the assumptions laid out only at the true $\psi_j$, in general. As a result, $U_j^*(\psi_j)$ form unbiased estimating equations which are uniquely solved at the true $\psi_j$, and as a result produce consistent estimators for $\psi_j$.
\end{proof}

\begin{proof}[Theorem~\ref{thm::modified_g_distribution}]
    Under non-exceptional laws, and the standard regularity conditions, then Theorem~\ref{thm::modified_G_consistency} demonstrates that the $U^*$ is an unbiased estimating equation. Supposing that the other nuisance parameters are estimated via M-estimation techniques, then a simple invocation of two-step M-estimation theory provides the necessary results.    
\end{proof}

\section{Additional Simulation Results}
In this section we continue the first simulation setting, reporting the results for the other parameter values. Quoting directly from the setup in the main article, \begin{quote}
    [...] we consider a two-stage DTR with two independent tailoring variates $X_1 \sim N(1,1)$ and $X_2 \sim N(1,4)$. We take $\pr(A_j^* = 1 \mid X_j) = \expit(X_j)$, for $j=1,2$, where $\expit(x) = (1+\exp(-x))^{-1}$ is the inverse logistic function. Given assigned treatment and $X_j$, we specify $\pr(A_j = 1\mid A_j^*, X_j) = \expit(-4.6 - 0.83X_j + 7.5A_j^*)$. The parameter values result in fairly low misclassification rates with values of approximately $0.01$ and $0.05$ for those prescribed $A_j^* = 0$ and $A_j^* = 1$, respectively. At stage one the contrast function is $1 + X_1$ and at stage two it is $1 + X_2 + \psi_{22}A_1$, where $\psi_{22}$ is varied across the grid $\{-1,-0.1,0,0.1,1\}$. The treatment-free component is $X_1$, and the outcome follows a normal distribution with variance $2$. Data are simulated with a sample size of $1000$, using an internal validation sample of $30\%$. The simulations are repeated $1000$ times [...]. Estimation compares an as treated analysis using standard G-estimation where the true treatment variables are available, standard G-estimation using the prescribed treatment in place of truth, modified G-estimation with estimated nonadherence probabilities, and modified G-estimation assuming the nonadherence rates were known.
\end{quote}

\begin{figure}
    \centering
    \includegraphics[width=\textwidth]{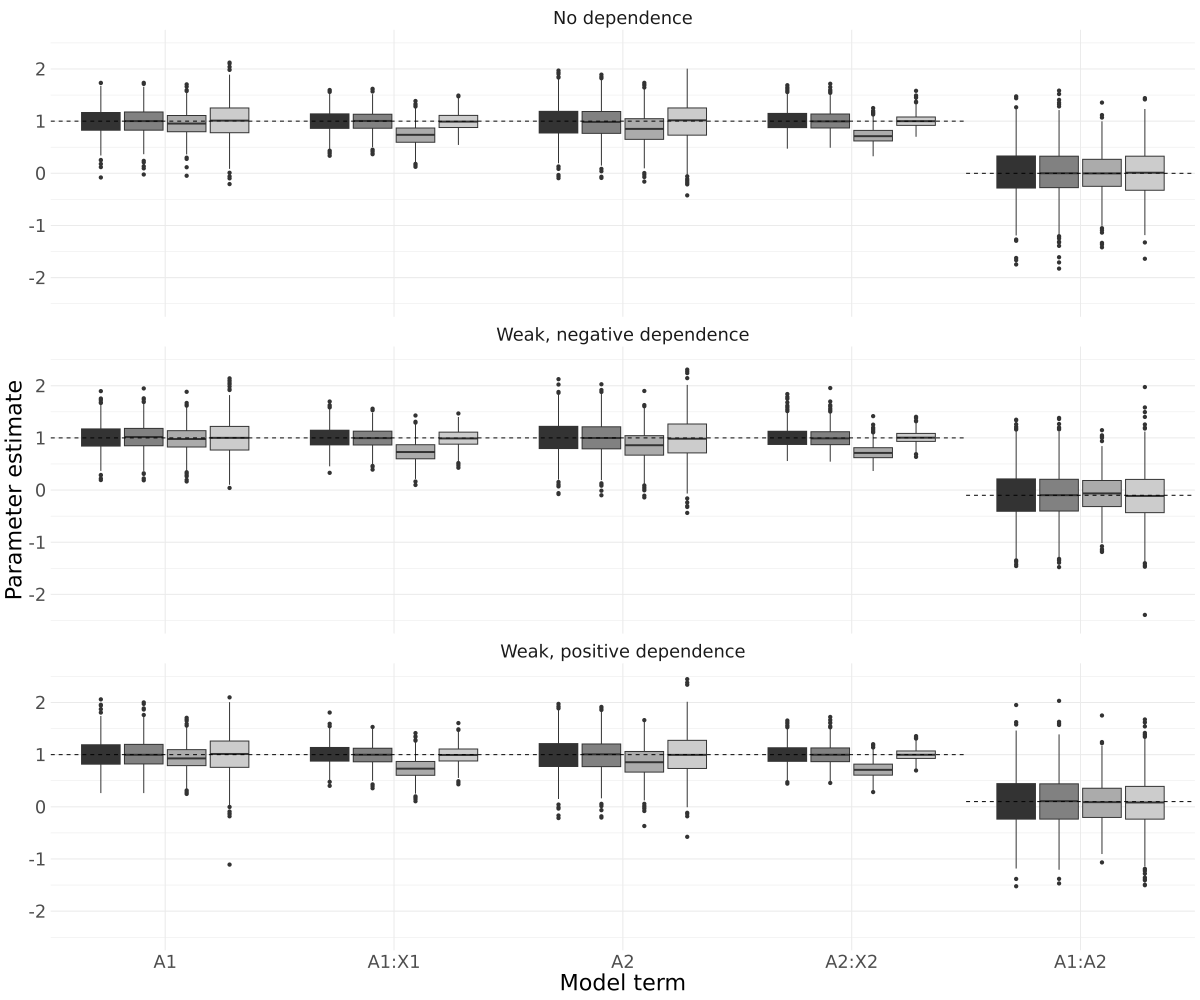}
    \caption{Estimated parameter values for a two-stage dynamic treatment regime comparing $\psi_{22} = 0$ (No dependence), $\psi_{22} = -0.1$ (weak, negative dependence), and $\psi_{22} = 0.1$ (weak, positive dependence). For each model term, from left to right, we compare the results of the modified G-estimation with known adherence rates (darkest), the modified G-estimation with estimated adherence rates (moderately dark), the standard G-estimation using prescribed treatment (moderately light), and the standard G-estimation using the truth (lightest).}
    \label{fig::simulation_1_appdx}
\end{figure}

Figure~\ref{fig::simulation_1_appdx} includes the results for $\psi_{22} \in \{-0.1,0,0.1\}$, plotted as box plots across the three scenarios. The patterns observed in these simulation runs are similar to those displayed in the main article.

\end{document}